\documentclass[aps,preprint,prd,nofootinbib]{revtex4}

\usepackage{amsmath}
\usepackage{graphicx}
\usepackage[colorlinks,linkcolor=magenta,anchorcolor=cyan,citecolor=blue]{hyperref}

\def\be{\begin{equation}}
\def\ee{\end{equation}}
\def\ba{\begin{eqnarray}}
\def\ea{\end{eqnarray}}

\def\lf{\left}
\def\rt{\right}

\begin{document}

\title{Merger rate of supermassive primordial black hole binaries}

\author{Hai-Long Huang$^{1}$\footnote{huanghailong18@mails.ucas.ac.cn}}
\author{Jun-Qian Jiang$^{1}$\footnote{jiangjq2000@gmail.com}}
%\author{Jun Zhang$^{2,3}$\footnote{zhangjun@ucas.ac.cn}}
\author{Yun-Song Piao$^{1,2,3,4}$\footnote{yspiao@ucas.ac.cn}}

\affiliation{$^1$ School of Physical Sciences, University of
Chinese Academy of Sciences, Beijing 100049, China}

\affiliation{$^2$ International Center for Theoretical Physics
    Asia-Pacific, Beijing/Hangzhou, China}

%\affiliation{$^3$ Taiji Laboratory for Gravitational Wave Universe (Beijing/Hangzhou), University of Chinese Academy of Sciences, 100049 Beijing, China}

\affiliation{$^3$ School of Fundamental Physics and Mathematical
    Sciences, Hangzhou Institute for Advanced Study, UCAS, Hangzhou
    310024, China}

\affiliation{$^4$ Institute of Theoretical Physics, Chinese
    Academy of Sciences, P.O. Box 2735, Beijing 100190, China}

%\author{author$^{1}$\footnote{email@example.com}}

%\affiliation{$^1$ some affiliations}

\begin{abstract}

The probability that the primordial black hole (PBH) binaries
formed in the early Universe can be affected by the Hubble
expansion of background, which is non-negligible when the number
density of PBHs is very low (it is actually this case for
supermassive PBHs). In this paper, taking into account the effect
of cosmic expansion on the comoving distance of PBH pairs, we
worked out the merger rate of PBHs with
any extended mass function. The torques by all PBHs and linear
density perturbations are also considered. It is found that the
merger rate of PBH, $M\gtrsim 10^6M_\odot$, binaries is
significantly lower for $f_\text{pbh}\lesssim 0.01$ than expected.

\end{abstract}

\maketitle
%\tableofcontents

\section{Introduction}

Recently, the detection by LIGO/Virgo of the gravitational waves
(GWs) emitted by the merging of black hole (BH) binaries, $M\sim
10M_\odot$ (e.g.
\cite{LIGOScientific:2016aoc,LIGOScientific:2016sjg}) has revived
the interest in the models of PBHs
\cite{Zeldovich:1967lct,Hawking:1971ei,Carr:1974nx}, although
other explanations for the origins of such BHs as well as the
LIGO/Virgo binaries are also possible (see e.g. Ref.
\cite{LIGOScientific:2016vpg}). It has been widely thought that
PBHs can constitute a fraction of or all dark matter
\cite{Bird:2016dcv,Clesse:2016vqa,Sasaki:2016jop}.

The nano-Hertz stochastic GW background (SGWB) detected recently
by PTA experiments
\cite{NANOGrav:2023gor,Xu:2023wog,Reardon:2023gzh,EPTA:2023fyk}
might be interpreted with a population of supermassive BH binaries
with $M\gtrsim10^9M_\odot$ \cite{NANOGrav:2023hfp,Ellis:2023dgf}.
The developing space-based detectors, e.g. LISA
\cite{LISA:2017pwj} and Taiji \cite{Hu:2017mde}, which aim for mHz GWs,
are ideal for detecting the mergers of massive and
supermassive BHs with masses above $10^3M_\odot$,
e.g.\cite{LISACosmologyWorkingGroup:2023njw}. Supermassive BHs
also may have primordial inflationary origin (e.g.
\cite{Huang:2023chx}, see also
\cite{Hooper:2023nnl,Nakama:2016kfq}), i.e. supermassive PBHs
(SMPBHs). It is well-known that the result of merger rate of PBHs
is necessary for assessing the event rate of PBH mergers at
different redshift, which might be significant for distinguishing
PBHs from astrophysical BHs, and SGWB.

There are some earlier works on the merger rate of PBHs, which
were effectively randomly distributed in space and formed in the
early Universe \footnote{Conversely, see e.g.
\cite{Raidal:2017mfl,Desjacques:2018wuu,Inman:2019wvr,
DeLuca:2020jug} for an initially clustered spatial distribution
and \cite{Bird:2016dcv,Nishikawa:2017chy,Ali-Haimoud:2017rtz} for
the merger rate in the late Universe.}. %Based on Ref.\cite{Ioka:1998nz},
%showed that PBHs would form binaries in the early Universe.
The merger rate of PBHs in Refs.\cite{Sasaki:2016jop,
Sasaki:2018dmp} is applicable to the case that all PBHs have the
same (monochromatic) mass and the torque is exerted only by the
nearest PBH. In Ref.\cite{Ali-Haimoud:2017rtz}, the tidal torquing
by all other PBHs, as well as standard large-scale adiabatic
perturbations, has been also taken into account, see also
\cite{Chen:2018czv,Liu:2018ess} for the extended mass function.
Refs.\cite{Kocsis:2017yty,Raidal:2018bbj} also presented the
merger rate of PBHs with non-monochromatic mass. Though N-body
simulations have been performed in Ref.\cite{Raidal:2018bbj}, the
mass range of PBHs simulated is $\sim\mathcal{O}(10)M_\odot$.

However, when the number density of PBHs is very low (it is
actually this case for SMPBHs), the effect of cosmic Hubble
expansion on binding PBH binaries (specifically, on the comoving
separation of PBH pair\footnote{The physical distance of PBH pair
at the decoupling time is equal to the semi-major axis of the
resultant binary \cite{Kocsis:2017yty}, this can be converted to
the maximum for the semi-major axis as did in
Ref.\cite{Sasaki:2018dmp}.}), so the merger rate, might be not
negligible. Relevant studies,
e.g.Refs.\cite{Ali-Haimoud:2017rtz,Chen:2018czv,Liu:2018ess}, did
not take this effect into account. The current observations
require the fraction of PBHs $f_\text{pbh}\lesssim 10^{-3}$ for
$10^6M_\odot\lesssim M\lesssim 10^{12}M_\odot$, thus the number
density $\sim \rho_\text{pbh}M^{-1}$ is actually considerably low
for SMPBHs.

In this paper, taking into account the effect of cosmic expansion
on comoving separation of PBH pairs, we show our merger rate in
section-\autoref{sec:mergerate}, and discuss its implications in
section-\autoref{sec:effect}. Throughout this paper we use units
$c=G=1$ and the values of cosmological parameters are set in light
of the Planck results \cite{Planck:2018vyg}. We denote by $t_0$
the present time. The scale factor is
normalized to unity at the matter radiation equality
$z=z_\text{eq} \approx3400$.

\section{Merger rate}
\label{sec:mergerate}

The probability distribution function of PBHs is
e.g.\cite{Raidal:2018bbj,Hall:2020daa,Franciolini:2022tfm}
\begin{equation}
    \psi(m)\equiv\frac{m}{\rho_\text{pbh}}\frac{\text{d}n}{\text{d}m},
\end{equation}
with $\psi(m)$ normalized as $\int\psi(m)\text{d}m=1$, where
$\rho_\text{pbh}$ is the energy density of PBHs, $n(m)$ is the
average number density of PBHs in the mass interval
$(m,m+\text{d}m)$.

The comoving total average number density of PBHs is
\begin{equation}
n_T=\frac{\rho_\text{pbh}}{\lf\langle m\rt\rangle},
\end{equation}
where $\lf\langle m\rt\rangle\equiv\frac{1}{n}\int
m\text{d}n=\lf(\int\text{d}m\frac{\psi(m)} {m}\rt)^{-1}$ is the
mean mass of PBHs with $n=\int\text{d}n(m)$. The fraction of the
average number density of PBHs in the total average number density
of PBHs is \cite{Liu:2018ess,Liu:2019rnx}
\begin{equation}
    F(m)\equiv\frac{n(m)}{n_T}=\psi(m)\frac{\lf\langle m\rt\rangle}{m},
\end{equation}
and the characteristic comoving separation between nearest PBHs
with $m_i$ and $m_j$ is \cite{Ali-Haimoud:2017rtz}
\begin{equation}
    \Bar{x}=\lf(\frac{3}{4\pi n_T}\rt)^{1/3}.
\end{equation}

The matter density at matter-radiation equality is
\begin{equation}
\rho_{\text{eq}}=\rho_{m,0}(1+z_{\text{eq}})^3,
\end{equation}
where $\rho_{m,0}$ is the matter density at present. Then the
comoving energy density of PBH is
$\rho_\text{pbh}=f\rho_{\text{eq}}$, with
$f\equiv\rho_\text{pbh}/\rho_m\approx 0.85f_\text{pbh}$, while
$f_\text{pbh}\equiv\rho_\text{pbh}/\rho_\text{dm}$ is the fraction
of PBHs in dark matter. The energy density of radiation in the
radiation era is
\begin{equation}
    \rho_r(z)=\rho_\text{eq}\lf(\frac{1+z}{1+z_\text{eq}}\rt)^4.
\end{equation}
The condition for two PBHs to become gravitationally bound is that
the total energy $m_i+m_j$ of PBHs must exceed the background
energy contained in the comoving bulk to the nearest PBH
\cite{Kocsis:2017yty}, i.e.
\begin{equation}
    m_i + m_j > \frac{8\pi}{3}\lf(x\cdot\frac{1+z_\text{eq}}{1+z}\rt)^3\rho(z),
\end{equation}
where $x$ is the comoving separation (i.e. the proper separation
at $z_\text{eq}$) of the PBH pair. It is expected that only when
%$x$ is smaller than $x_\text{max}$ given by
\begin{equation}
    x<x_\text{max}=\lf(\frac{3}{8\pi}\cdot\frac{m_i+m_j}{\rho_\text{eq}}\rt)^{1/3},
\end{equation}
can the PBH pair come into being at redshift
$z=z_\text{dec}>z_\text{eq}$ with
\begin{equation}
    1+z_\text{dec}=(1+z_\text{eq})\lf(\frac{x_\text{max}}{x}\rt)^{3}.
\end{equation}

By solving the equation of motion for the proper separation
projected along the axis of motion of PBH pair numerically, one
can get the semi-major axis $a$ of the binary
\cite{Ali-Haimoud:2017rtz}
\begin{equation} \label{eq:a}
    a\approx0.1\lambda x,
\end{equation}
with
\begin{equation}
    \lambda=\frac{8\pi\rho_\text{eq}x^3}{3(m_i+m_j)}=\lf(\frac{x}{x_\text{max}}\rt)^3<1.
\end{equation}
\autoref{eq:a} also shows that the semi-major axis is equivalent
to the physical distance of the PBH pair at the decoupling time
numerically, i.e. $a\sim x(1+z_\text{eq})/(1+z_\text{dec})$
\cite{Kocsis:2017yty}.

In the case of sparse PBH distribution, the condition
$x<x_\text{max}$ (or equivalently, $\lambda<1$) must be taken into
account in the probability distribution $P(m_i,m_j,x)$ of $x$
between two nearest PBHs with masses $m_i$ and $m_j$ (but without
other PBHs in the bulk of $4\pi x^3/3$). Assuming that PBHs
possess a random distribution, we have
\begin{align}\label{eq:dPdX}
\text{d}P(m_i,m_j,X)&=4\pi x^2\Theta(x_\text{max}-x)\text{d}x\left[F(m_i)\text{d}m_i
n(m_j)\text{d}m_j e^{-4\pi x^3n(m_j)\text{d}m\over 3}\prod_{m\neq
m_j}e^{-4\pi x^3n(m)\text{d}m\over 3}\right]
\notag \\
&=\Theta(X_\text{max}-X)\text{d}X
F(m_i)\text{d}m_iF(m_j)\text{d}m_je^{-X},
\end{align}
where $X\equiv(x/\Bar{x})^3={4\pi n_Tx^3/ 3}$, with
\begin{equation} \label{eq:Xmax}
    X_\text{max}\equiv\lf(\frac{x_\text{max}}{\Bar{x}}\rt)^3=\frac{(m_i+m_j)n_T}
    {2\rho_\text{eq}}.
\end{equation}

Then we need to calculate the dimensionless angular momentum $j$
of the PBH pair with the torque by all surrounding PBHs and
density perturbations of the rest dark matter. The probability
distribution of $j$ for a given $X$ is
\begin{equation} \label{eq:jdPdj}
    j\lf.\frac{\text{d}P}{\text{d}j}\rt|_X=\mathcal{P}(j/j_X), \quad
    \mathcal{P}(\gamma)\equiv\frac{\gamma^2}{(1+\gamma^2)^{3/2}},
\end{equation}
see e.g. Refs.\cite{Ali-Haimoud:2017rtz,Liu:2018ess,Chen:2018czv}
for detailed derivation, where the characteristic value of $j_X$
is estimated as \cite{Liu:2018ess}
\begin{equation} \label{eq:jX}
    j_X\approx\frac{\lf\langle m\rt\rangle}{m_i+m_j}(1+\sigma_\text{eq}^2/f^2)^{1/2}X.
\end{equation}
Here $\sigma_\text{eq}^2$ is the variance of density perturbations
of the rest of dark matter at $z_\text{eq}$
\cite{Ali-Haimoud:2017rtz}. As a comparison, the angular momentum
only accounting for tidal torquing by the nearest PBH with mass of
$m_l$ (the comoving distance is $x'$) is estimated as
\cite{Nakamura:1997sm,Ioka:1998nz}
\begin{equation}\label{eq:jX1997}
    j\sim \frac{2m_l}{m_i+m_j}\lf(\frac{x}{x'}\rt)^3,
\end{equation}
which is physically natural only for a (or nearly) monochromatic
mass function. However, since the torque is also proportional to
the mass of the outer PBH, \autoref{eq:jX1997} is not valid if the
mass of PBHs extends over many orders of magnitude. Thus we adopt
\autoref{eq:jdPdj}.

The angular momentum $j$ can be expressed by the semi-major axis
$a$ and the coalescence time $t$ as \cite{Peters:1964zz}
\begin{equation} \label{eq:j}
    j=\lf[\frac{85}{3}\cdot\frac{m_im_j(m_i+m_j)}{a^4}\rt]^{1/7}t^{1/7}.
\end{equation}
Then we have
\begin{equation}
    \gamma_X=\frac{j(t;X)}{j_X}=Ct^{1/7}X^{-37/21}.
\end{equation}
The factor $C$ only depends on $m_i$ and $m_j$. Combining
\autoref{eq:dPdX}, \autoref{eq:jdPdj}, \autoref{eq:jX} and
\autoref{eq:j}, we can get the probability distribution of the
merger time
\begin{align} \label{eq:lastdpdt}
    \frac{\text{d}P(m_i,m_j,X)}{\text{d}t}&=\int\text{d}X\frac{\text{d}^2P(m_i,m_j,X)}{\text{d}X\text{d}t}
\notag \\
    &=\frac{1}{7t}F(m_i)\text{d}m_iF(m_j)\text{d}m_j\int\text{d}Xe^{-X}
    \Theta(X_\text{max}-X)\mathcal{P}(\gamma_X).
\end{align}
%We also show in \autoref{sec:XstarandXmax} the relationship between the values of
%$X_*(m_i,m_j,t)$ and 1, $X_\text{max}(m_i,m_j)$.
The peak of $\mathcal{P}(\gamma_X)$ is at $X_*\ll 1$, see
Appendix-\autoref{sec:XstarandXmax}, which suggests that
$e^{-X}\approx1$ in \autoref{eq:lastdpdt}
\cite{Ali-Haimoud:2017rtz,Chen:2018czv,Liu:2018ess}. Thus with the
generalized hypergeometric function ${}_2F_1$, the integral in
\autoref{eq:lastdpdt} can be integrated out as
\begin{align} \label{eq:intXmaxP}
    \int\text{d}X\Theta(X_\text{max}-X)\mathcal{P}(\gamma_X)&=\frac{21X_\text{max}^{58/21}}
    {2146(C^2t^{2/7}+X_\text{max}^{74/21})Ct^{1/7}}\lf[95C^2t^{2/7}{}_2F_1\lf(-\frac{1}
    {2},\frac{29}{37},\frac{66}{37},-\frac{X_\text{max}^{74/21}}{C^2t^{2/7}}\rt)\rt.
    \notag \\
    &-(58C^2t^{2/7}+21X_\text{max}^{74/21})\lf.{}_2F_1\lf(\frac{1}
    {2},\frac{29}{37},\frac{66}{37},-\frac{X_\text{max}^{74/21}}{C^2t^{2/7}}\rt)\rt],
\end{align}
It is convenient to rewrite \autoref{eq:intXmaxP} as
$\int\text{d}X\Theta(X_\text{max}-X)\mathcal{P}(\gamma_X)=X_\text{max}Y(y)$
with
\begin{equation} \label{eq:hy}
    Y(y)=\frac{21y}{2146(1+y^2)}\lf[95{}_2F_1\lf(-\frac{1}{2},\frac{29}{37},
    \frac{66}{37},-y^2\rt)-(58+21y^2){}_2F_1\lf(\frac{1}{2},\frac{29}{37},
    \frac{66}{37},-y^2\rt)\rt],
\end{equation}
where
\begin{align} \label{eq:y}
    y(m_i,m_j,t)&\equiv\frac{1}{\gamma_X(t;X_\text{max})}=\frac{X_\text{max}^{37/21}}{Ct^{1/7}}
    \notag \\
    &\approx
    2.95\times10^2f\lf(1+\frac{\sigma_\text{eq}^2}{f^2}\rt)^{\frac{1}{2}}
    \lf(\frac{m_i}{M_\odot}\rt)^{-\frac{1}{7}}\lf(\frac{m_j}{M_\odot}\rt)^{-\frac{1}{7}}
    \lf(\frac{m_i+m_j}{M_\odot}\rt)^{\frac{1}{21}}
    \lf(\frac{t}{t_0}\rt)^{-\frac{1}{7}}.
\end{align}

Thus our resulting merger rate is
\begin{equation}
    R(t)=\frac{\text{d}N_\text{merge}}{\text{d}t\text{d}V}=\frac{1}{2}\frac{n_T}{(1+z_
    \text{eq})^3}\frac{\text{d}P}{\text{d}t}
    \equiv\int\int\mathcal{R}(m_i,m_j,t)\text{d}m_i\text{d}m_j,
\end{equation}
where $\mathcal{R}(m_i,m_j,t)$ is the differential merger rate,
\begin{align} \label{eq:mergerate}
    \mathcal{R}(m_i,m_j,t)&\approx\frac{1.02\times10^8}{\text{Gpc}^3\text{yr}}f^2
    \lf(\frac{m_i}{M_\odot}\rt)^{-1}\lf(\frac{m_j}{M_\odot}\rt)^{-1}\lf(\frac{m_i+m_j}
    {M_\odot}\rt)\lf(\frac{t}{t_0}\rt)^{-1}
    \notag \\
    &\times Y\lf(y(m_i,m_j,t)\rt)\psi(m_i)\psi(m_j).
\end{align}

\section{Implication of our merger rate}
\label{sec:effect}

\begin{figure}[h!]
\centering
\includegraphics[width=0.65\textwidth]{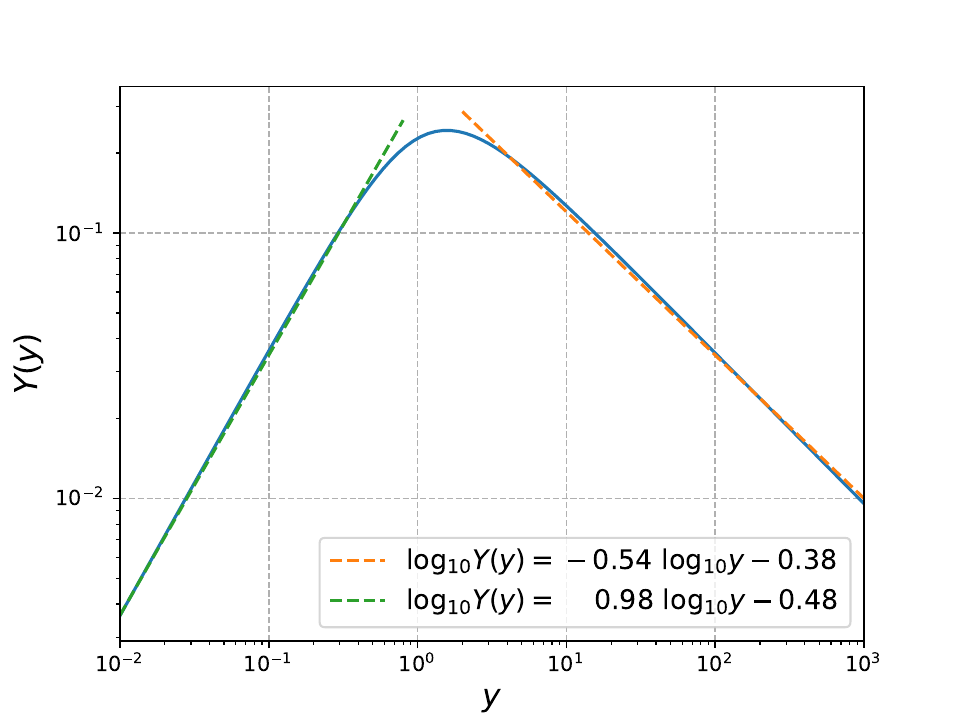}
\caption{\label{fig:Hy} $Y(y)$ with respect to $y$ according to
\autoref{eq:hy}. }
\end{figure}

In the condition that the number density of PBHs is large,
resulting in $y\ge1$ according to \autoref{eq:y} and
$Y(y)\approx0.42y^{-0.54}$ approximately, see \autoref{fig:Hy}.
After substituting $Y(y)$ into \autoref{eq:mergerate} combined
with \autoref{eq:y}, we arrive that
\begin{align}\label{eq:largenumber}
    \mathcal{R}(m_i,m_j,t)&\approx\frac{1.99\times10^6}{\text{Gpc}^3\text{yr}}f^{1.46}
    \lf(1+\frac{\sigma_{\text{eq}}^2}{f^2}\rt)^{-0.27}
    \lf(\frac{m_i}{M_\odot}\rt)^{-0.92}\lf(\frac{m_j}{M_\odot}\rt)^{-0.92}
    \notag \\ & \times
    \lf(\frac{m_i+m_j}{M_\odot}\rt)^{0.97}
    \lf(\frac{t}{t_0}\rt)^{-0.92}\psi(m_i)\psi(m_j).
\end{align}
\autoref{eq:largenumber} corresponds to the merger rate without
the limitation $X<X_\text{max}$, i.e. the effect of cosmic
expansion on the comoving distance of PBH pairs is negligible.
In e.g. Ref.\cite{Liu:2018ess}, such a
merger rate density has been showed as
\begin{align} \label{eq:liulang}
    \mathcal{{R}}_{no-expan}(m_i,m_j,t)&=\frac{1.94\times10^6}{\text{Gpc}^3\text{yr}}
    f^{\frac{53}{37}}
    \lf(1+\frac{\sigma_{\text{eq}}^2}{f^2}\rt)^{-\frac{21}{74}}
    \lf(\frac{m_i}{M_\odot}\rt)^{-\frac{34}{37}}\lf(\frac{m_j}{M_\odot}\rt)^{-\frac{34}{37}}
    \notag \\
    &\times
    \lf(\frac{m_i+m_j}{M_\odot}\rt)^{\frac{36}{37}}
    \lf(\frac{t}{t_0}\rt)^{-\frac{34}{37}}\psi(m_i)\psi(m_j),
\end{align}
which is essentially consistent with our \autoref{eq:largenumber}.

\begin{figure}[h!]
\centering
\includegraphics[width=0.75\textwidth]{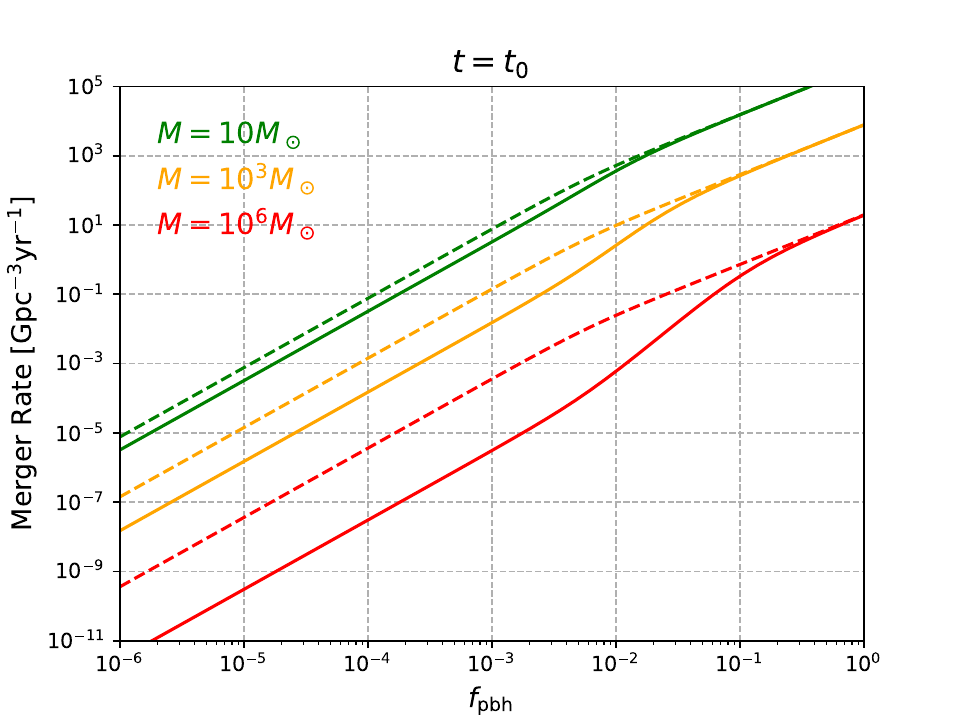}
\caption{\label{fig:mono} Merger rate $R(t_0)$ at present of
monochromatic PBH binaries with masses $10M_\odot$ (green),
$10^3M_\odot$ (orange) and $10^6M_\odot$ (red), respectively. As a
comparison, we use different merger rate formulas, i.e.
\autoref{eq:mergerate} (our work, solid lines) and
\autoref{eq:liulang} (dashed lines). }
\end{figure}

Taking the monochromatic PBHs
\begin{equation} \label{eq:monoc}
    \psi(m)=\delta(m-M),
\end{equation}
we plot $R(t_0)$ and ${R}_{no-expan}(t_0)$ for different $M=10,
10^3, 10^6M_\odot$ in \autoref{fig:mono}, respectively. As
expected, we always have $R(t)\lesssim{R}_{no-expan}(t)$. And the
smaller $M$ or the larger $f_\text{pbh}$, the closer both are.
This suggests that the effect of cosmic expansion on the
separation can be safely neglected only if the number density of
PBHs $\sim \rho_\text{pbh}M^{-1}$ is very high.

\begin{figure}[h!]
\centering
\includegraphics[width=1\textwidth]{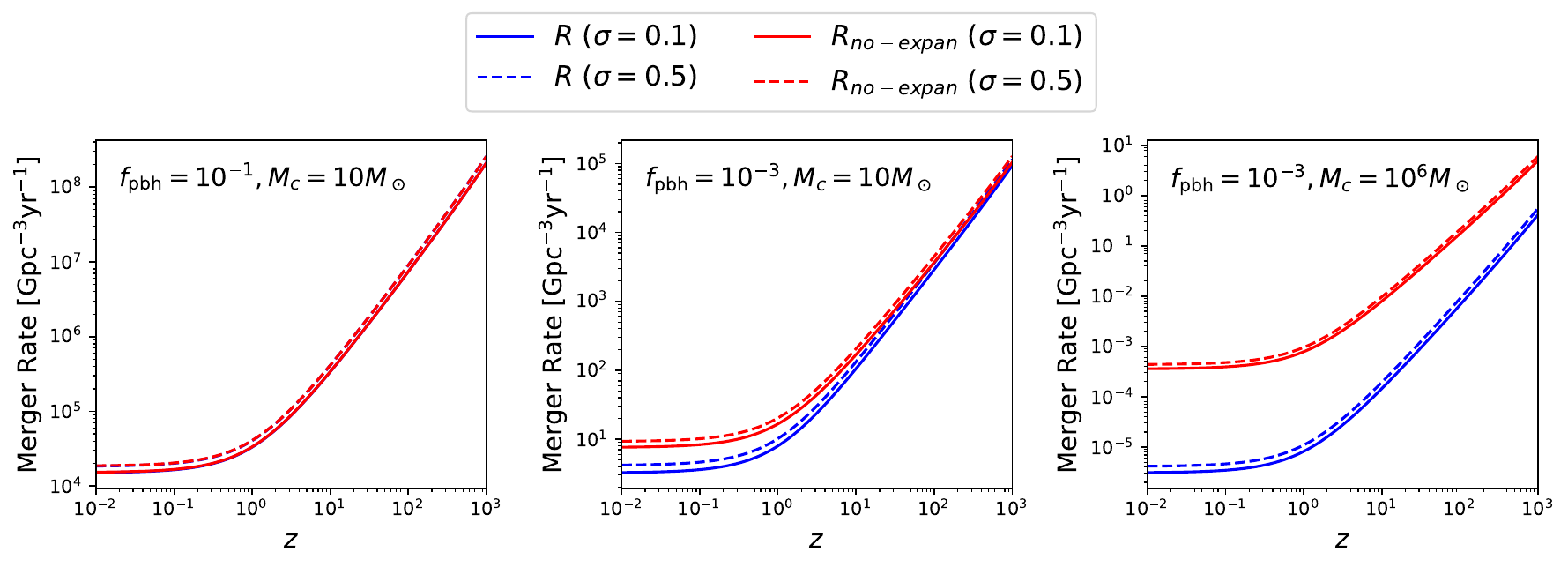}
\caption{\label{fig:mergerate_redshift} Merger rate of PBHs
(sourced by supercritical bubbles) with an extended mass function
\autoref{eq:bubbleMFa2} with respect to the redshift according to
\autoref{eq:mergerate} (blue, our work) and \autoref{eq:liulang}
(red, Ref. \cite{Liu:2018ess}). }
\end{figure}

It is also interesting to consider that of PBHs sourced by
supercritical bubbles that nucleated during slow-roll inflation,
see Appendix-\autoref{sec:bubbleMassFunction},
\begin{equation} \label{eq:bubbleMFa2}
    \psi(m)=e^{-\sigma^2/8}\sqrt{\frac{M_c}{2\pi\sigma^2 m^3}}\exp
    \lf(-\frac{\ln^2(m/M_c)}{2\sigma^2}\rt).
\end{equation}
In \autoref{fig:mergerate_redshift}, we plot $R(t)$ and
${R}_{no-expan}(t)$ at different redshifts, respectively. It is
clearly seen that when $f_\text{pbh}=10^{-1}$ and the
characteristic mass $M_c=10M_\odot$, the difference between $R(t)$
and ${R}_{no-expan}(t)$ is indistinguishable, while when
$f_\text{pbh}\sim 10^{-3}$, the difference becomes apparent,
especially at low redshifts. This is because the later binary
merge corresponds to the larger separation $x$, so more
non-negligible the effect of cosmic expansion. Thus the effect of
cosmic expansion will be significant in suppressing the merger
rate for SMPBHs.

\section{Conclusion}

In the early Universe the probability that the PBHs formed might
be affected by the Hubble expansion of background, which is
non-negligible when the spatial distribution of PBHs is sparse,
specially for SMPBHs. In this paper, taking the effect of cosmic
expansion on the comoving separation of PBH pair into account, we
worked out the merger rate of PBHs with the extended mass
function, see \autoref{eq:mergerate}.
%The torques by all other PBHs and the
%linear density perturbations have been taken into account.
It is found that the merger rate of SMPBHs, $M\gtrsim
10^6M_\odot$, can be significantly suppressed for
$f_\text{pbh}\lesssim 0.01$\footnote{Current cosmological and
astrophysical constraints for the abundance of SMPBHs are
$f_\text{pbh}\lesssim10^{-3}$ \cite{Carr:2023tpt}.}.

Throughout our estimation, the subdominant effects, such as the
effects of the tidal field from the smooth halo, the encountering
with other PBHs, the baryon accretion, present-day halos and the
spin of PBHs, are neglected. In order to study the corresponding
effects thoroughly, it might be better to perform the N-body
simulation in an expanding background. Nonetheless, for the merger
of PBHs with any extended mass function, our
\autoref{eq:mergerate} might suffice to capture the essential
impact of cosmic expansion on the merger rate, which thus can have
interesting applications, specially for SMPBHs and low-frequency
GWs.

\section*{Acknowledgments}
We thanks Yong Cai, Lang Liu and Jun Zhang for helpful discussion.
YSP is supported by NSFC, No.12075246, National Key Research and
Development Program of China, No. 2021YFC2203004, and the
Fundamental Research Funds for the Central Universities.

\appendix

\section{On $X_*$ and $X_{\text{max}}$}
\label{sec:XstarandXmax}

The value of $X_*$ maximizes $\mathcal{P}(\gamma_X)$, i.e.
\begin{equation}
    \mathcal{P}'(\gamma_{X_*})\lf.\frac{\partial\gamma_X}{\partial X}\rt|_{X=X_*}=0.
\end{equation}
Since $\gamma_X$ is monotonic, which implies
$\mathcal{P}'(\gamma_{X_*})=0$, we get
\begin{equation}\label{eq:jtXstar}
    j(t;X_*)=\sqrt{2}j_{X_*},
\end{equation}
According to \autoref{eq:jtXstar}, we have
\begin{equation} \label{eq:Xstar}
    X_*
    \approx1.63\times10^{-2}f^{\frac{16}{37}}\lf(1+\frac{\sigma_{\text{eq}}^2}{f^2}\rt)^
    {-\frac{21}{74}}\lf(\frac{m_i}{M_\odot}\rt)^{\frac{3}{37}}\lf(\frac{m_j}{M_\odot}\rt)^{\frac{3}{37}}\lf(\frac{m_i+m_j}{M_\odot}\rt)^{\frac{36}{37}}
    \lf(\frac{\lf\langle m\rt\rangle}{M_\odot}\rt)^{-1}\lf(\frac{t}{t_0}\rt)^{\frac{3}{37}}.
\end{equation}

%$X_*$ as $\mathcal{P}(\gamma_X)$ has a sharp peak at

\begin{figure}[h!]
\centering
\includegraphics[width=0.75\textwidth]{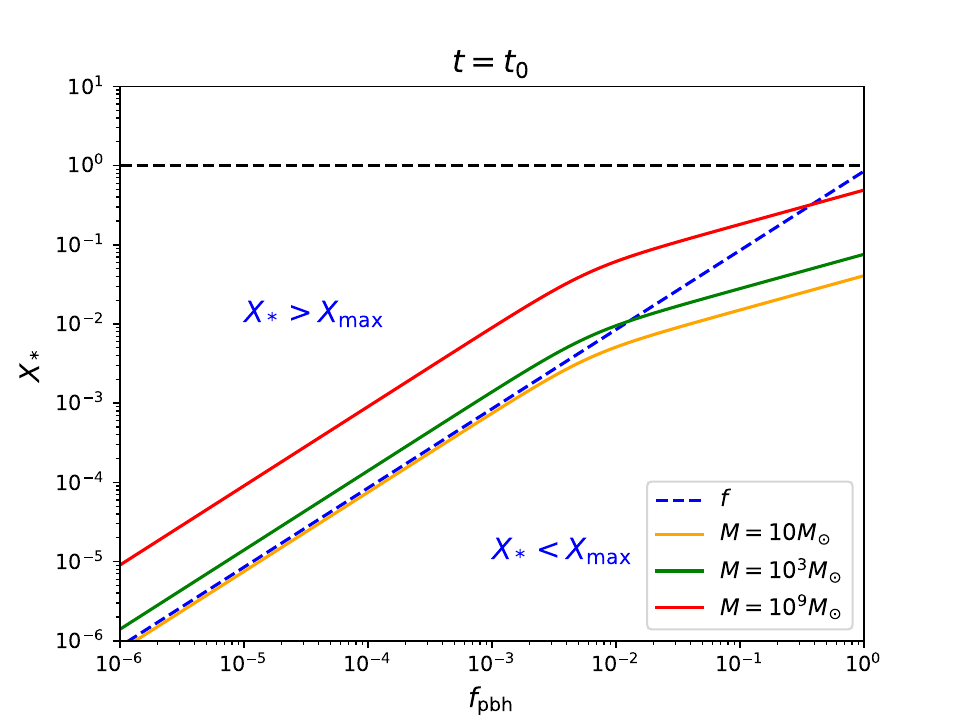}
\caption{\label{fig:XstarwithXmax} The most probable value of $X$
for binaries merging today for the PBH population with
monochromatic mass function. One can compare it with Fig. 2 in
\cite{Ali-Haimoud:2017rtz}. The blue dotted line represents
$X_\text{max}=f$, above which is the parameter space satisfying
$X_*>X_\text{max}$. We can see that $X_*\ll1$ still holds in the
case of SMPBHs, considering the cosmological and astrophysical
constraints on the abundance for SMPBHs
\cite{Carr:2020gox,Carr:2018rid} (e.g. we have
$f_\text{pbh}<10^{-3}$ for $M=10^9M_\odot$). Thus it is safe to
neglect the factor $e^{-X}$ in \autoref{eq:lastdpdt}. However, we
do not have $X_*\ll X_\text{max}$ similarly, especially for
SMPBHs.
%Where the slope changes depends on the value of
Here, $\sigma_\text{eq}=0.005$. }
\end{figure}

In the monochromatic mass approximation, \autoref{eq:monoc}, we
have $X_\text{max}=f$. Thus
\begin{equation}
    X_*\approx3.20\times10^{-2}f^{\frac{16}{37}}\lf(1+\frac{\sigma_{\text{eq}}^2}{f^2}\rt)^
    {-\frac{21}{74}}\lf(\frac{M}{M_\odot}\rt)^{\frac{5}{37}}\lf(\frac{t}{t_0}\rt)^{\frac{3}{37}}.
\end{equation}
We plot $X_*$ compared with 1 and $X_\text{max}=f$ in
\autoref{fig:XstarwithXmax}, which shows that $X_*\ll1$ in all
cases while $X_*>X_\text{max}$ for SMPBHs.

\section{Mass distribution of PBHs sourced by supercritical bubbles}
\label{sec:bubbleMassFunction}

The bubbles that nucleated during slow-roll inflation can
naturally develop to the PBHs \cite{Garriga:2015fdk,Deng:2016vzb}.
It is found in Ref.\cite{Huang:2023chx}\footnote{Here, the
nucleating rate of bubbles must satisfy $\lambda\ll 1$. In the
case of $\lambda\gtrsim 1$, the resulting scenario corresponds to
a two-stage inflation model with an intermediate first-order phase
transition, during which the collisions between bubbles that
nucleated will contribute inflationary GWB, e.g.
\cite{Wang:2018caj,Li:2020cjj}. } that the PBHs sourced by
supercritical bubbles not only can be supermassive, $M\gtrsim
10^{9}M_\odot$, but also have a peak-like mass
function\footnote{See also Ref.\cite{Kleban:2023ugf} for a
different perspective. }
\begin{equation} \label{eq:bubbleMF}
    \psi(m)\sim\frac{(c_1c_2^3)^{\frac{1}{2}}}{f_{\text{pbh}}M_
    {\odot}}\exp\lf\{-B_*\lf[1+c_3\lf(\mathcal{N}^{1/2}-\mathcal
    {N}_*^{1/2}\rt)^2\rt]^{n/2}-3{\cal N}\rt\},
\end{equation}
with $c_1=\mathcal{M}_{\mathrm{eq}}/M_\odot\approx10^{17}$,
$c_2=H_iM_\odot/ M_p^2\approx10^{32}$ and
$\mathcal{N}=\ln\sqrt{c_2(m/M_\odot)}$, where $n=1$ for the
nucleation of \textsf{domain wall}
\cite{Garriga:2015fdk,Basu:1991ig} and $n=4$ for the
\textsf{vacuum bubble} \cite{Coleman:1980aw}, $H_i$ is the
inflationary Hubble parameter, $\mathcal{N}$ is the efolds number
before inflation ended. In corresponding model, we naturally have
the parameters $c_3\gg1$, $B_*\sim\mathcal{O}(10)$ and
$\mathcal{N}_*\sim\mathcal{O}(10)$.

\begin{figure}[h!]
\centering
\includegraphics[width=0.55\textwidth]{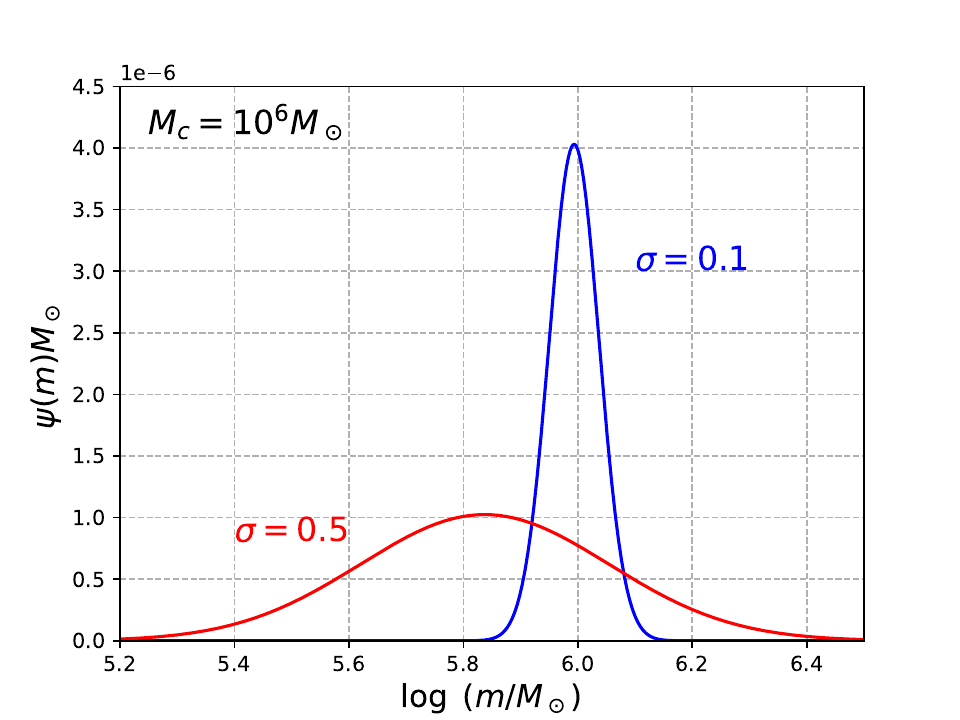}
\caption{\label{fig:massfunction} The normalized mass function of
PBHs sourced by supercritical bubbles. }
\end{figure}

It is convenient to convert $\mathcal{N}_*$ to the characteristic
mass $M_c$ by
\begin{equation}
    \mathcal{N}_*\equiv\ln\sqrt{c_2(M_c/M_\odot)}.
\end{equation}
%\autoref{eq:bubbleMF} is a unimodal distribution. We can take
In the approximation of $\mathcal{N}\to\mathcal{N_*}$ (or
equivalently, $m\to M_c$), we have
\begin{equation}
    \psi(m)\propto\frac{1}{m^{3/2}}\exp\lf\{-B_*\lf[1+c_3\frac{\ln^2(m/M_c)}
    {16{\cal N}_*}\rt]^{n/2}\rt\}\sim\frac{1}{m^{3/2}}\exp\lf[-B_*\lf(1+\frac{\ln^2(m/M_c)}{2\sigma^2}\rt)\rt],
\end{equation}
where $\sigma\equiv4\sqrt{\mathcal{N}_*/(nc_3)}$ corresponds to
the width of mass peak. Thus the normalized mass function is
\begin{equation} \label{eq:bubbleMFa}
    \psi(m)=e^{-\sigma^2/8}\sqrt{\frac{M_c}{2\pi\sigma^2 m^3}}\exp
    \lf(-\frac{\ln^2(m/M_c)}{2\sigma^2}\rt).
\end{equation}
It is obvious that $\psi(m)$ approaches the monochromatic spectrum
centered on $M_c$ as $\sigma\to0$, see \autoref{fig:massfunction}.
According to \autoref{eq:bubbleMFa}, we have
\begin{equation}
    \lf\langle m \rt\rangle=M_ce^{-\sigma^2}, \quad
    \text{and} \quad
    \lf\langle m^2 \rt\rangle=M_c^2e^{-\sigma^2}.
\end{equation}

\bibliography{refs}

\end{document}